\documentclass[aps,prl,showpacs,preprintnumbers,twocolumn]{revtex4}
\usepackage{epsfig,graphicx}

\input{tcilatex}

\begin{document}

\preprint{CAS-KITPC/ITP-004}

\title{Designing a Spin Channel for Perfect Quantum Communication}
\author{Xiao-Qiang Xi$^{1,\ 2}$, T. Zhang$^{3}$, R. H. Yue$^{3,\ 4}$, X. C. Xie$^2$, W. M. Liu$^2$}
\address{$^1$Department of Applied Mathematics and Physics,
 Xi'an Institute of Posts and Telecommunications, Xi'an 710061, China}
\address{$^2$Beijing National Laboratory for Condensed Matter Physics,
Institute of Physics, Chinese Academy of Sciences, Beijing 100080,
China}
\address{$^3$Institute of Modern Physics, Northwest
University, Xi'an 710069, China}
\address{$^4$Department of Physics, Ningbo
University, Ningbo 315211, China}

\date{\today}

\begin{abstract}
We propose a scheme for using a spin chain with fixed symmetry
interaction as a channel for perfect quantum communication. The
perfect quantum communication is determined by the eigenvalues that
form a special arithmetical progression, the concrete interaction
parameter $J_i$ is obtained as a function of integer and perfect
transmission time $t_p$. There are infinite choices of $J_i$ for
perfect transmission and one can design the channel according to the
requirement of $t_p$. This scheme will provide more choices of spin
chain for future experiment in quantum communication.
\end{abstract}


\maketitle {\emph{Introduction.}} In quantum information processing,
it is needed to transfer a quantum state from one place to another,
such as quantum key distribution \cite{GisinRMP74_145},
teleportation \cite{BennettPRL70_1895} and quantum computation
(communication between quantum processors) \cite{BennettN404_247}.
Thus it is very important to find physical systems to serve as
channels for quantum communication. A very successful area is
quantum optics, in which the implementation of quantum state
transmission has been realized. The carriers of information can be
addressed and transmitted with high control and with a low level of
decoherence, while this kind of state transport needs interfacing
between quantum processor and quantum channels, which make the
quantum computer complexity and lower speed.

So a great attention is focused on the problem of transferring
quantum information in a solid-state environment, in which one can
either first use the channel to share entanglement with a separated
party and then use this entanglement for teleportation, or can
directly transmit a state through the channel. The key step of using
teleportation to transport a state is looking for a quantum channel
with ideal entanglement, such as a spin chain with ideal boundary
entanglement \cite{XiAPS55_3026,VenutiPRL96_247206,Xi0609087}.

Bose first propose a scheme to use a spin chain as a channel for
short distance quantum communication \cite{BosePRL_207901}. The
communication is achieved by state dynamical evolution in spin
chain, which does not require the ability to switch ``on'' and
``off'' the interactions between the spins, it also does not require
any modulation by external fields either. This approach is
restricted to zero temperature and single spin-wave states. Basing
on Bose's scheme, many works focus on the perfect transmission in
spin systems \cite{
LiPRA71_022301,ChristandlPRL92_187902,VerstraetePRL92_087201,
OsbornePRA69_052315,BurgarthPRA71_052315,
AlbanesePRL93_230502,KarbachPRA72_030301}. Christandl {\em et al.}
\cite{ChristandlPRL92_187902} found the simplest scheme to make a
proper choice of the modulation of the coupling strengths in spin
chain then realize a perfect transmission in arbitrary long
distance, Zhang {\em et al.} \cite{ZhangPRA72_012331} realize this
scheme in three qubit case by using liquid nuclear magnetic
resonance. Seeking for the ideal channel is important, as well as
its transmitted time, the shorter the better. Christandl's scheme
\cite{ChristandlPRL92_187902} is masterly, but the interaction
$J_i=J_{N-i}=\sqrt{i(N-i)}$ for perfect transmission is not unique.

In this paper, the shortest time of perfect transmission will be
studied in spin chain, as well as the corresponding properties of
interaction by constructing the eigenvectors with the ways different
from Ref. \cite{BosePRL_207901,ChristandlPRL92_187902}.

\emph{The eigenvectors and the transmit amplitude that described the
transmission.} The Hamiltonian of N qubit Heisenberg XXX open chain
is $H_N^{xxx}= \sum_{i=1}^{N-1} [J_i(\sigma_{i}^{+} \sigma_{i+1}^{-}
+\sigma_{i+1}^{+} \sigma_{i}^{-})+\frac{J_i}{2}\sigma_{i}^{z}
\sigma_{i+1}^{z}]$, where $\sigma^{\pm}= \frac{1}{2} (\sigma^x \pm
i\sigma^y)$, $\sigma^{x}, \sigma^{y}, \sigma^{z}$ are the Pauli
matrices, $J_i$ the exchange hopping. If the $\sigma^z$ term
vanishes, this system degenerates into XX model note as
$H_N^{xx}=\sum_{i=1}^{N-1} J_i(\sigma_{i}^{+} \sigma_{i+1}^{-}
+\sigma_{i+1}^{+} \sigma_{i}^{-})$.

 The initial state of the system is prepared as the first qubit to be spin
up and the others spin down. Since the Hamiltonian commutes with the
total spin component along the $z$ direction, the relevant sector of
the Hilbert space must be spanned by the states
$|j>=|0_1,0_2,\dots,0_{j-1},1_j,0_{j+1},\dots,0_N>$ with
$j=1,\dots,N$.

The open spin chain is symmetrical about the middle point. Let
$J_i=J_{N-i}$, $|j>$ and $|N+1-j>$ must have the same probability in
the eigenvectors, so the eigenvectors of the system can be written
as {\footnotesize
\begin{eqnarray}
|\psi_m>&=&\sum_{j=1}^{k}C_{mj}(e^{i\alpha_{mj}}|j>+|N+1-j>), N=2k;
\nonumber \\
|\psi_m>&=&\sum_{j=1}^{k}C_{mj}(e^{i\alpha_{mj}}|j>+|N+1-j>);
\nonumber
\\
& &+C_{m,k+1}|k+1>, N=2k+1
\end{eqnarray}}
where $C_{mj},\ \alpha_{mj}$(0 or $\pi),\ j=1,2,\dots,k,
m=1,2,\dots,N$ are the parameters determined by $H|\psi>=E|\psi>$
and the normalization condition.

The transition amplitude between the boundary qubits is
\begin{equation}
f_{1N}^{Nxx(x)}(J_i,t)=\sum_{m=1}^N<N|\psi_m><\psi_m|1>e^{-iE_mt}.
\end{equation}

\emph{Four qubit XXX open chain and finite qubits XX open chain as
ideal channel.} In four qubit chain, let $J_1=J_3=1$ and $J_2=J$.
$J=1$ corresponds the uniform chain. The eigenvalues and the
corresponding eigenvectors are shown in Table 1.

{\footnotesize Table 1. The eigenvalues and eigenvectors of four
qubit XXX open chain.}
\begin{center}
{\footnotesize
\begin{tabular}{c|c|c|c}
\hline m & $E_m$ &$C_{m1}(\alpha_{m1})$ & $C_{m2}(\alpha_{m2})$ \cr
\hline

1 & $\frac{J}{2}+1$ & $\frac{1}{2}(0)$ &$\frac{1}{2}(0)$ \cr\hline

2 & $\frac{J}{2}-1$ & $\frac{1}{2}(0)$ &$-\frac{1}{2}(0)$ \cr \hline

3 & $-\frac{J}{2}+\sqrt{J^2+1}$
&$\sqrt{\frac{1}{2+2(\sqrt{J^2+1}-J)^2}}(\pi)$
&$\sqrt{\frac{(\sqrt{J^2+1}-J)^2}{2+2(\sqrt{J^2+1}-J)^2}}(\pi)$ \cr
\hline

4 & $-\frac{J}{2}-\sqrt{J^2+1}$
&$\sqrt{\frac{1}{2+2(\sqrt{J^2+1}+J)^2}}(\pi)$
&$-\sqrt{\frac{(\sqrt{J^2+1}+J)^2}{2+2(\sqrt{J^2+1}+J)^2}}(\pi)$ \cr
\hline
\end{tabular}}
\end{center}

Using Eq. (2) one can get the transition amplitude between the
boundary qubits:
$f_{14}^{4xxx}(J,t)=C_{11}^2e^{-iE_1t}+C_{21}^2e^{-iE_2t}
-C_{31}^2e^{-iE_3t}-C_{41}^2e^{-iE_4t}$. In order to obtain perfect
transition (i.e. $|f_{14}^{4xxx}|=1$), two conditions must be
satisfied at the same time, they are $\sum_{m=1}^4C_{m1}^2=1$ and
$e^{-iE_1t}=e^{-iE_2t}=-e^{-iE_3t}=-e^{-iE_4t}=1$ (or $i$). The
first condition is satisfied naturally, but the second not, the
reason is that $\frac{E_1}{E_2}=\frac{J+2}{J-2}=\frac{k_1}{k_2}$,
$\frac{E_3}{E_4}=\frac{J-2\sqrt{J^2+1}}{J+2\sqrt{J^2+1}}=\frac{2k_3+1}{2k_4+1}$
and $\frac{J+2}{J-2\sqrt{J^2+1}}=\frac{2k_1}{2k_3+1}$ can not be
satisfied exactly at the same time, where $k_i\ (i=1,2,3,4)$ is
integer. So this model can not give perfect transmission.

In Ref. \cite{BosePRL_207901}, Bose manipulated a uniform open XXX
chain with magnetic field which has no effect to the transition
amplitude, and claimed that ``$N=4$'' gives perfect quantum
transmission, his result based on ``numerical analysis''. The
eigenvectors Eq. (9) in Ref. \cite{BosePRL_207901} are the same as
Table 1. when J=1, while the eigenvalues are just different in a
constant which comes from the magnetic field, the corresponding
order of eigenvectors are $|\tilde{1}>-|\psi_1>$,
$|\tilde{2}>-|\psi_3>$, $|\tilde{3}>-|\psi_2>$ and
$|\tilde{4}>-|\psi_4>$. Although four qubit XXX open chain can not
realize perfect transmission, proper value $J$ in Table 1 can make
$|f_{14}^{4xxx}|\rightarrow 1$, see Table 2.

{\footnotesize Table 2. The relation between the maximal transition
amplitude and $J$.}

\begin{center}
{\footnotesize
\begin{tabular}{c c c||c c c}
\hline J & t & $|f_{14}^{4xxx}|_{max}$& J & t &
$|f_{14}^{4xxx}|_{max}$\cr \hline
 1& 29$\pi$& 0.99981 & 7& 14$\pi$& 0.99997\cr \hline
 2& 17$\pi$& 0.99978 & 8& 16$\pi$& 0.99998\cr \hline
 3& 6$\pi$& 0.99914  & 9& 18$\pi$& 0.99999\cr \hline
 4& 8$\pi$& 0.99970  & 10& 20$\pi$& 0.99999\cr \hline
 5& 10$\pi$& 0.99988 & $\dots$& $\dots$ & $>$0.99999\cr \hline
 6& 12$\pi$& 0.99994 & N& 2N$\pi$& $>$0.99999 \cr \hline
\end{tabular}}
\end{center}

In order to get perfect transmission, we must simplify the
restricted conditions. The natural idea is to simplify the model, so
we change XXX model as XX model and calculate its transition
amplitude
$|f_{14}^{4xx}(J,t)|=|-i(2C_{11}^2\sin{(E_1t)}+2C_{21}^2\sin{(E_2t)})|$,
here $C_{11}=\sqrt{1/(2+2E_1^2)},\ C_{21}=\sqrt{1/(2+2E_2^2)}$,
$E_{1,2}=\frac{J\pm\sqrt{J^2+4}}{2}$. Perfect transmission means
$|f_{14}^{4xx}|=1$, as in the XXX model, two conditions must be
satisfied in the XX model: (1) $2C_{11}^2+2C_{21}^2=1$; (2)
$\frac{E_1}{E_2}
 =\frac{J+\sqrt{J^2+4}}{J-\sqrt{J^2+4}}=\frac{2k_1+1}{2k_2+1}$,
 $k_1,\ k_2$ are integer and $k_1-k_2$ must be even number. The first condition
is satisfied for any $J$, the second condition is satisfied when
$J=\sqrt{\frac{4(k_1+k_2+1)^2}{(k_1-k_2)^2-(k_1+k_2+1)^2}}$, perfect
transmission can be realized at time
$t=\frac{(2k_1+1)\pi}{2E_1(J)}$. When $k_1=1$ and $k_2=-1$,
$J={2}/{\sqrt{3}}$, i.e. the interaction in the middle is stronger
than the boundary, at this condition, the time of perfect
transmission is $t_{min}=\frac{\sqrt{3}\pi}{2}$, this is coincidence
with $J_1=J_3=\sqrt{3}$ and $J_2=2$ in Ref.
\cite{ChristandlPRL92_187902}. When $k_1=2$ and $k_2=-2$,
$J={2}/{\sqrt{15}}$, i.e. the interaction in the middle is weaker
than the boundary, the time of perfect transmission is
$t_{min}=\frac{\sqrt{15}\pi}{2}$. If $k_1$ and $k_2$ take other
values, one can obtain corresponding interaction for perfect
transmission.

Although four qubit XX open chain can realize perfect transmission,
its length is not enough for a practical utilization, which request
us to looking for a longer channel. The transition amplitude of XXX
model in Table 2. show that the interaction coupling $J$ in the
middle qubit has great effect to the transmission. In the
Hamiltonian of six qubit XX open chain $H_6^{xx}$ with $J_1=J_5,\
J_2=J_4, J_3$, the eigenvectors are
$|\psi_m>=\sum_{j=1}^{3}C_{mj}(e^{i\alpha_{mj}}|j>+|7-j>)$,
$m=1,2,\dots,6$. Using $H|\psi_m>=E_m|\psi_m>$ and the normalization
condition one can obtain $E_m$ ($E_1=-E_6,E_2=-E_5,E_3=-E_4$) and
the corresponding $C_{mj}$ and $\alpha_{mj}$. The transmit amplitude
between the boundary qubits is
$|f_{16}^{6xx}(J_i,t)|=|-i(2C_{11}^2\sin(E_1t)+2C_{21}^2\sin(E_2t)
+2C_{31}^2\sin(E_3t))|$. The aim of us is getting perfect
transmission, i.e. $|f_{16}^{6xx}(J_i,t)|=1$. The analysis shows
that $2C_{11}^2+2C_{21}^2+2C_{31}^2=1$, so the condition
$\sin(E_1t)=\sin(E_2t)=\sin(E_3t)=1$ (or $-1$) can assure perfect
transmission. Supposing $t_p$ is the time of perfect transmission,
then $E_1=\frac{(1+4n)\pi}{2t_p},\ E_2=\frac{\pi}{2t_p}$ and
$E_3=\frac{(1-4n)\pi}{2t_p}$, $n=1,2,\dots,$ can make
$\sin(E_1t)=\sin(E_2t)=\sin(E_3t)=1$. Using these conditions the
value of $J_1,\ J_2$ and $J_3$ can be solved, they are
{\footnotesize
\begin{equation}
J_1=J_5=\sqrt{\frac{(16n^2-1)\pi^2}{12t_p^2}},
J_2=J_4=\sqrt{\frac{(8n^2-2)\pi^2}{3t_p^2}}, J_3=\frac{3\pi}{2t_p}.
\end{equation}}
When $n=1$ and $t_p=\frac{\pi}{2}$, the $J_i$ in Eq. (3) satisfy
$J_i=J_{5-i}=\sqrt{i(5-i)}$ \cite{ChristandlPRL92_187902}. If n
takes other integer, one can obtain the corresponding perfect
transmission interaction $J_i$. For example if $n=2,\
t_p=\frac{\pi}{2}$ in Eq. (3) one will obtain $J_1=J_5=\sqrt{31},\
J_2=J_4=\sqrt{40}$ and $J_3=3$, the interactions array an order
``weak-strong-weak-strong-$\dots-$'', maybe this kind of interaction
can be realized easily than $J_i=J_{N-i}=\sqrt{i(N-i)}$ which
increase monotonously from the boundary to middle.

In eight qubit XX open chain $H_8^{xx}$ with $J_1=J_7,\ J_2=J_6,\
J_3=J_5$ and $J_4$, the number of eigenvalue is eight, the
eigenvalues satisfy $E_i=-E_{8-i}, i=1,2,3,4$. The transmit
amplitude between the boundary qubits is
$|f_{18}^{8xx}(J_i,t)|=|-i(2C_{11}^2\sin(E_1t)+2C_{21}^2\sin(E_2t)
+2C_{31}^2\sin(E_3t)+2C_{41}^2\sin(E_4t))|$. As in six qubit case,
perfect transmission is assured by the condition
$\sin(E_1t)=\sin(E_2t)=\sin(E_3t)=\sin(E_4t)=1$ (or $-1$). Supposing
$t_p$ is the time of perfect transmission, then
$E_1=\frac{(8n-1)\pi}{2t_p},\ E_2=\frac{(4n-1)\pi}{2t_p},\
E_3=-\frac{\pi}{2t_p}$
 and $E_4=-\frac{(4n+1)\pi}{2t_p}$ with
$n=1,2,\dots,$ can ensure perfect transmission. Using these
eigenvalues the concrete expression of $J_i$ can be solved, they are
{\footnotesize\begin{eqnarray}
J_1&=&J_7=\frac{\pi}{2t_p}\sqrt{\frac{32n^2+4n-1}{5}},\nonumber \\
J_2&=&J_6=
\frac{\pi}{2t_p}\sqrt{\frac{48n^2+16n-4}{5}}, \nonumber \\
 J_3&=&J_5=\frac{\pi}{2t_p}\sqrt{20n-5},\  J_4=\frac{(8n-4)\pi}{2t_p}
\end{eqnarray}}
The $J_i$ corresponds to the results in
\cite{ChristandlPRL92_187902} if $n=1$, except this, $J_i$ has
infinite choices for perfect transmission. For example $n=2,\
t_p=\frac{\pi}{2}$ in Eq. (4) one will obtain $J_1=J_7=\sqrt{27},\
J_2=J_6=\sqrt{44}$ and $J_3=J_5=\sqrt{35}$ and $J_4=12$, these is an
order ``weak-strong-weak-strong-$\dots-$''.

In odd qubit open chain, three qubit case is a uniform chain,
$J_1=J_2$, the transition amplitude between the boundary qubits is
$|f_{13}^{3xx}(t)|=\frac{1}{2}|(\cos{\sqrt{2}J_1t}-1)|$, perfect
state transfer can be obtained at $t=\frac{\pi}{\sqrt{2}J_1}$. For
five qubit case, $J_1=J_4,\ J_2=J_3$,
$|f_{15}^{5xx}|=|\frac{J_2^2}{2J_2^2+J_1^2}+\frac{J_1^2}{2(2J_2^2+J_1^2)}\cos{(\sqrt{2J_2^2+J_1^2}
t)} -\frac{1}{2}\cos{J_1t}|$. When $J_1=\frac{\pi}{t_p},\
J_2=\sqrt{\frac{(2k)^2-1}{2}}\frac{\pi}{t_p},\ k=1,2,3,\dots$ or
$J_2\gg{J_1}$, perfect transmission can be obtained at
$t_p=\frac{\pi}{J_1}$. For seven qubit case, we can get
{\footnotesize
\begin{eqnarray}
J_1&=&J_6=\frac{\pi}{t_p}\frac{(2n_1-1)(2n_2-1)}{\sqrt{A}},\nonumber \\
J_2&=&J_5=\frac{\pi}{t_p}\frac{\sqrt{(2n_3)^2A-(2n_1-1)^2(2n_2-1)^2}}
{\sqrt{A}},\nonumber \\
 J_3&=&J_4=\frac{\pi}{t_p}\sqrt{\frac{A}{2}},
\end{eqnarray}} where $A=(2n_1-1)^2+(2n_2-1)^2-(2n_3)^2$, $n_1,n_2,n_3=1,2,\dots,$
and the value of $n_i,\ i=1,2,3$ must keep $J_i\ i=1,2,3$ real.
Similarly as in even qubit case, one has infinite choices for
perfect transmission except for $J_i=J_{N-i}=\sqrt{i(N-i)}$. The
different lies in one need more than one integer to determine $J_i$.

{\emph{N qubit XX open chain as ideal channel.}} For general N qubit
case, the idea of constructing ideal channel is the same as finite
qubit case: at the condition of the spin chain with symmetry
interaction $J_i=J_{N-i}$, one can figure out the expression of
transmit amplitude $|f_{1N}^{Nxx}|$, then find the perfect
transmission is determined by the eigenvalues of the system, at use
the proper eigenvalues to solve the corresponding $J_i$. The
transmit amplitude expression has great difference for the parity of
N.

(1) N is even, at this case, one have to divided it into two classes
because there still has some difference between $N=4k$ and $N=4k-2$,
$k=1,2,3,\dots$.

When $N=4k$, the transmit amplitude at this case can be written as
$|f_{1N}^{Nxx}|=|-i(\sum_{j=1}^{2k}2C_{j1}^2\sin E_jt)|$, the number
of $E_j>0$ and $E_j<0$ is equal, as long as $E_jt$ can construct a
special arithmetical progression $a_i\
(a_0=(-1-4(k-1)n)\frac{\pi}{2},a_{i+1}-a_i=(4n)\frac{\pi}{2}\
n=1,2,\dots)$, then we can realize perfect transmission. The
concrete expression of $J_i$ is obtained by comparing the
coefficients in the following two equations
{\footnotesize
\begin{eqnarray}
&&\prod_{j=1}^{2k}
(x-E_j)=\prod_{i=-(k-1)}^{k}[x-(-1+4in)\frac{\pi}{2t}]=0 \\
&&Det\left(\begin{array}{ccccc}
-x&J_1&0&\dots&0 \\
J_1&-x&J_2&\dots&0 \\
0&J_2&-x&\dots&0 \\
\vdots&\vdots&\vdots&\ddots&J_{2k-1} \\
0&0&\dots&J_{2k-1}&J_{2k}-x\\
 \end{array}\right)_{2k\times 2k}=0,
\end{eqnarray}}

When $N=4k-2$, the transmit amplitude is
$|f_{1N}^{Nxx}|=|-i(\sum_{j=1}^{2k-1}2C_{j1}^2\sin E_jt)|$, the
number of $E_j>0$ is $k$ and the number of $E_j<0$ is $k-1$, as long
as $E_jt$ can construct a special arithmetical progression $a_i\
(a_0=(1-4(k-1)n)\frac{\pi}{2},a_{i+1}-a_i=(4n)\frac{\pi}{2})$, then
we can realize perfect transmission. $J_i$ is obtained by comparing
the coefficients in the following two equations
{\footnotesize
\begin{eqnarray}
&&\prod_{j=1}^{2k}
(x-E_j)=\prod_{i=-(k-1)}^{k-1}[x-(1+4in)\frac{\pi}{2t}]=0,
\\
&&Det\left(\begin{array}{ccccc}
-x&J_1&0&\dots&0 \\
J_1&-x&J_2&\dots&0 \\
0&J_2&-x&\dots&0 \\
\vdots&\vdots&\vdots&\ddots&J_{2k-2} \\
0&0&\dots&J_{2k-2}&J_{2k-1}-x\\
 \end{array}\right)_{(2k-1)\times (2k-1)}=0. \nonumber \\
\end{eqnarray}}

(2) N is odd, we still divided it into two classes because the
slightly difference between $N=4k-1$ and $N=4k+1$, where
$k=1,2,3,\dots$.

When $N=4k-1$, the transmit amplitude is
$|f_{1N}^{Nxx}|=|\sum_{j=1}^{k}2C_{j1}^2\cos
E_jt-(C_{2k+k,1}^2+\sum_{j=2k+1}^{2k+k-1}2C_{j1}^2\cos E_jt)|$, the
number of $E_j>0$ and $E_j<0$ is equal, the number of $E_j=0$ is
one. $J_i$ is obtained by comparing the following two group
equations,
{\footnotesize
\begin{eqnarray}
&&\prod_{j=1}^{k} (x-E_j)=\prod_{j=1}^{k}(x-\frac{2n_j\pi}{t})=0,
\\
&&Det\left(\begin{array}{ccccc}
-x&J_1&0&\dots&0 \\
J_1&-x&J_2&\dots&0 \\
0&J_2&-x&\dots&0 \\
\vdots&\vdots&\vdots&\ddots&J_{2k-1} \\
0&0&\dots&2J_{2k-1}& -x\\
 \end{array}\right)_{2k\times 2k}=0,
\\
&&x\prod_{j=2k+1}^{2k+k-1}
(x-E_j)=x\prod_{j=2k+1}^{2k+k-1}(x-\frac{(2n_j-1)\pi}{t})=0,
\\
&&Det\left(\begin{array}{ccccc}
-x&J_1&0&\dots&0 \\
J_1&-x&J_2&\dots&0 \\
0&J_2&-x&\dots&0 \\
\vdots&\vdots&\vdots&\ddots&J_{2k-2} \\
0&0&\dots&J_{2k-2}& -x\\
 \end{array}\right)_{(2k-1)\times (2k-1)}=0, \nonumber \\
\end{eqnarray}}
where $n_j=1,2,\dots$ and their values must satisfy $J_i$ is real,
similarly in $N=4k+1$ case.

When $N=4k+1$, the transmit amplitude is
$|f_{1N}^{Nxx}|=|-(C_{k+1,1}^2+\sum_{j=1}^k2C_{j1}^2\cos
E_jt)+\sum_{j=2k+2}^{2k+k+1}2C_{j1}^2\cos E_jt|$, the number of
$E_j>0$ and $E_j<0$ is equal, the number of $E_j=0$ is one. $J_i$ is
obtained by comparing the following two group equations,
{\footnotesize
\begin{eqnarray}
&& x\prod_{j=1}^{k} (x-E_j)=\prod_{j=1}^{k}(x-\frac{2n_j\pi}{t})=0,
\\
&& Det\left(\begin{array}{ccccc}
-x&J_1&0&\dots&0 \\
J_1&-x&J_2&\dots&0 \\
0&J_2&-x&\dots&0 \\
\vdots&\vdots&\vdots&\ddots&J_{2k} \\
0&0&\dots&2J_{2k}& -x\\
 \end{array}\right)_{(2k+1)\times (2k+1)}=0,
\\
&& \prod_{j=2k+2}^{2k+k+1}
(x-E_j)=\prod_{j=2k+2}^{2k+k+1}(x-\frac{(2n_j-1)\pi}{t})=0,
\\
&& Det\left(\begin{array}{ccccc}
-x&J_1&0&\dots&0 \\
J_1&-x&J_2&\dots&0 \\
0&J_2&-x&\dots&0 \\
\vdots&\vdots&\vdots&\ddots&J_{2k-1} \\
0&0&\dots&J_{2k-1}& -x\\
 \end{array}\right)_{2k\times 2k}=0.
\end{eqnarray}
}

Someone may doubt how can we solve such equations when $N$ is large
enough. The doubt is reasonable but here is unnecessary, because
there exist a recursion relation between $J_i$ and $J_{i+1}$, the
interaction in the middle is easily obtained. Obviously, even qubit
chain has some advantage than odd qubit chain, the former only need
one parameter $n$ to determine $J_i$, while the later need $(N-1)/2$
parameters $n_1,n_2,\dots,n_{\frac{N-1}{2}}$ to determine $J_i$,
from this point even qubit chain is a good candidate for experiment.

\emph{Discussions.} Our results offer infinite choices of
interaction $J_i$ for perfect transmission. The perfect transmission
time $t_p\propto {1}/{J_i}$, it will be short enough as long as
$J_i$ can be tuned to arbitrarily large. Our scheme can be
generalized to transmit more than one bit information. The further
calculation show that four qubit open chain with symmetry
interaction can be used to transmit two bits information perfectly.

It is also interesting to study the boundary entanglement of the
perfect transmission channel. For four qubit XXX open chain, the
boundary entanglement is {\footnotesize
\begin{equation}
C_{14}=\frac{2(C_{11}^2e^{-\frac{E_1}{T}}+C_{21}^2e^{-\frac{E_2}{T}}-C_{31}^2
e^{-\frac{E_{3}}{T}}-C_{41}^2e^{-\frac{E_{4}}{T}})}{\sum_{m=1}^4e^{-\frac{E_m}{T}}},
\end{equation}}
where $C_{i1}$ and $E_m$ can be seen from Table 1. The concurrence
$C_{14}$ is a half if $J\rightarrow 0$ and vanishes as J to be
enough large. The boundary entanglement of N qubit XX open chain is
$C_{1N}=\frac{1}{2^{N-2}}$ under the interaction
$J_i=\sqrt{i(N-i)}$. For other perfect interaction, one can also
calculate the corresponding $C_{1N}$, for example, N=4 XX open chain
with $J_1=J_3=1,\ J_2=\frac{2}{\sqrt{15}}$, $C_{14}=\frac{3}{8}$;
N=5 XX open chain with $J_1=J_4=1,\ J_2=J_3={\sqrt{30}}$,
$C_{15}=\frac{1}{16}$ etc.. Therefore we can conclude that the
perfect transmission channel is not the ideal entanglement channel.

In summary, we found a way of constructing an ideal spin channel for
quantum communication in study of the finite spin chain with
symmetry interaction $J_i=J_{N-1}$. The equation of $J_i$ satisfied
ideal channel condition and the corresponding solutions are
obtained. Besides possessing the advantage of Bose's and
Christandl's propose, our scheme offers infinite choices of
interaction for perfect quantum communication, and one can design an
ideal channel according to the requirement of the perfect
transmission time $t_p$. This scheme will provide more choices of
spin chain for future experiment in quantum communication and it can
also be generalized to transmit more than one bit information.

This research is supported in part by the Project of Knowledge
Innovation Program (PKIP) of Chinese Academy of Sciences, by the NSF
of China under grant 10547008, 90406017, 90403019, 60525417,
10610335, by the National Key Basic Research Special Foundation of
China under 2005CB724508, 2006CB921400,
 by the Foundation of Xi'an Institute of Posts
and Telecommunications under grant 105-0416.

\end{document}